\documentclass{raa_twocolumn}            

\usepackage{graphicx,times}             
\usepackage{natbib}
\usepackage{amssymb,amsmath}
\usepackage{listings}
\bibpunct{(}{)}{;}{a}{}{,}
\usepackage[pagebackref=true]{hyperref}

\begin{document}
	
	\title{Overview of the ECLAIRs Trigger for SVOM gamma-ray burst detection}
	
	\volnopage{Manuscript RAA-2026-0061 accepted for publication on 19-Feb-2026}      
	\setcounter{page}{1}          
	
	\author{S. Schanne 
		\inst{1,*}\footnotetext{$*$Corresponding Author.}, 
		F. Château \inst{2}, 
		N. Dagoneau\inst{2}, 
		F. Daly\inst{1},
		H. Le Provost\inst{2},
		and P. Kestener\inst{2} 
	}
	
	\institute{
		CEA Saclay/Irfu/DAp-AIM, CNRS, Université Paris-Cité, Université Paris-Saclay, 91191 Gif-sur-Yvette, France\\         \and
		CEA Saclay/Irfu/DEDIP, 91191 Gif-sur-Yvette, France\\
		\vs\no
		{\small Received 202x month day; accepted 202x month day}}
	
	\abstract{ 
		The French–Chinese SVOM satellite mission (Space-based multi-band astronomical Variable Objects Monitor) was launched in mid-2024, with science objectives focused on the detection and study of astrophysical transient events, primarily Gamma-Ray Bursts (GRBs). The onboard trigger of the hard X-ray wide-field coded-mask instrument ECLAIRs autonomously detects and localizes GRBs on SVOM and requests automatic spacecraft slews toward these sources, enabling follow-up observations by the onboard narrow field-of-view instruments MXT and VT. The trigger also transmits real-time alerts via the SVOM VHF network to the ground, allowing rapid follow-up campaigns by the broader community, including multiple space- and ground-based facilities.
		We present an overview of the ECLAIRs trigger, with emphasis on the two trigger algorithms running simultaneously onboard: the Count-Rate Trigger (CRT) and the Image Trigger (IMT), both of which issue alerts for GRBs localized on the sky. The trigger has already detected several notable events, including both classical GRBs and peculiar X-ray–rich GRBs, enabling numerous redshift measurements, including high-redshift bursts.
	}
	
	\authorrunning{S. Schanne et al
	}            
	\titlerunning{Overview of the ECLAIRs trigger onboard SVOM}  
	
	\maketitle
	
	%
	%
	
	\section{Introduction}
	
	SVOM (Space-based multi-band astronomical Variable Objects Monitor, \citealt{Wei+etal+2016}, \citealt{Cordier+etal+2026a}) is a space mission dedicated to the detection and study of Gamma-Ray Bursts (GRBs) and other astrophysical transients, providing real-time alerts to the worldwide community. 
	SVOM aims to build a well-characterized GRB sample with broad spectral and temporal coverage, including a high fraction of redshift measurements.
	SVOM was launched on June 22, 2024, into a 29° inclination low-Earth orbit. Following commissioning, nominal operations began early 2025, with a planned mission duration of at least five years.
	
	The GRB trigger of the ECLAIRs instrument onboard SVOM plays a central role in the SVOM system, where it initiates the whole GRB observation chain. It autonomously detects GRBs of various types and provides fast, reliable localizations with rapid alert dissemination. 
	
	The SVOM system comprises both space- and ground-based elements. 
	The onboard instruments for prompt GRB observations are:
	(i) ECLAIRs (\emph{Lightning} in French, \citealt{Godet+etal+2026a}), providing wide-field hard X-ray imaging (4–150 keV), triggering, localization, timing, and spectroscopy;
	(ii) GRM (Gamma-Ray Monitor, \citealt{Sun+etal+2026}) with three tilted detector modules, providing gamma-ray spectroscopy and timing (15 keV–5 MeV).
	Afterglow observations following an ECLAIRs-triggered slew are performed by:
	(iii) MXT (Microchannel X-ray Telescope, \citealt{Goetz+etal+2026}), refining the localization to $<$1 arcmin in X-rays;
	(iv) VT (Visible Telescope, \citealt{Qiu+etal+2026}), providing sub-arcsecond localization and optical light curves.
	Ground-based instruments include:
	(v) GWAC (Ground Wide-Angle Cameras, \citealt{Xin+etal+2026}), monitoring a large fraction of the ECLAIRs field of view for prompt optical emission;
	(vi) GFT (Ground Follow-up Telescopes, \citealt{Wu+etal+2026}, \citealt{Basa+etal+2026}) for optical and near-infrared observations.
	
	%
	The ECLAIRs trigger autonomously requests spacecraft slews towards detected GRBs, enabling automatic GRB afterglow follow-up by MXT and VT. VT then provides sub-arcsecond localizations, complemented by dedicated ground-based follow-up telescopes. Such precise localizations are essential for ground-based redshift measurements.
	SVOM generally points toward the night sky to favor rapid ground follow-up and redshift determination \citep{Cordier+etal+2008}. This strategy results in a periodically modulated background due to Earth passages through the ECLAIRs field of view, which the trigger handles robustly.
	
	Alerts generated by the ECLAIRs trigger are rapidly distributed to the ground via a VHF network \citep{Cordier+etal+2026b}, with redundancy provided by the Beidou system, and are processed at the French Science Center (FSC, \citealt{Louvin+etal+2026}), where alert notices are published for the follow-up community. The complete ECLAIRs data are stored onboard and downlinked via X-band for detailed offline analysis. Triggering is performed onboard, as the photon data volume is too large for real-time downlink.
	
	In addition to GRBs, the trigger supports broader transient science by detecting outbursts from known or unknown sources. These detections may prompt target-of-opportunity observations or follow-up of external multi-messenger alerts.
	
	\section{ECLAIRs instrument}
	
\begin{figure}
		\centering
		\includegraphics[width=0.5\textwidth, angle=0]{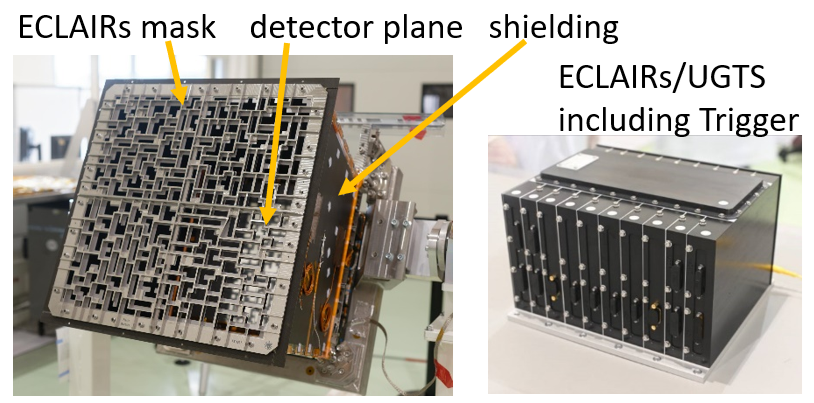}
		\caption{ECLAIRs flight model, basic parts (TXG: mask, detector, shielding; UGTS: onboard processing \& control).}
		\label{fig_eclairs-fm-pictures}
	\end{figure}

	ECLAIRs is a wide-field hard X-ray instrument (Fig. \ref{fig_eclairs-fm-pictures}) designed for real-time detection and localization of GRBs and other high-energy transients.
	Its low-energy detection threshold of 4 keV optimizes sensitivity to X-ray rich and high-redshift GRBs, while covering energies up to 120 keV for standard bursts. The 2 steradian field of view maximizes detection rates, and the localization accuracy of 12 arcmin at detection limit ensures source containment within the VT field of view after slew.
	ECLAIRs employs coded-mask imaging, based on expertise acquired from missions such as GRANAT, INTEGRAL, and Swift. The onboard trigger must generate slew requests within tens of seconds and communicate alerts rapidly to the ground. Due to limited telemetry in low-Earth orbit, all trigger analysis is performed by the scientific processing and control unit (UGTS) onboard.
	The instrument has an allocated mass of $\sim$90 kg and power consumption of $\sim$90 W.
	
	\subsection{Hard X-ray/soft gamma-ray telescope (TXG)}
	
	The TXG consists of a detection plane \citep{Godet+etal+2026b}, a coded mask \citep{Lachaud+etal+2026a}, and a passive shield defining the field of view.
	The detection plane comprises 6400 CdTe pixels arranged in an 80$\times$80 grid, covering $\sim$1000 cm$^2$. Pixels are read out by the sector electronics (ELS) and processed in real time by the UGTS.
	The 54$\times$54 cm$^2$ coded mask composed of Ta2.5W (Tantalum alloy with 2.5\% of Tungsten) has a 40\% open fraction and is optimized for imaging in the 4–120 keV range. Side shielding defines a $2$ sr field of view, with a $0.15$ sr fully coded region providing the highest sensitivity.
	Detector images (“shadowgrams”) are deconvolved onboard using the mask model to reconstruct sky images every few seconds, enabling rapid detection of new transient sources.
	
	\subsection{Scientific processing and control unit (UGTS)}
	
\begin{figure}
		\centering
		\includegraphics[width=0.5\textwidth, angle=0]{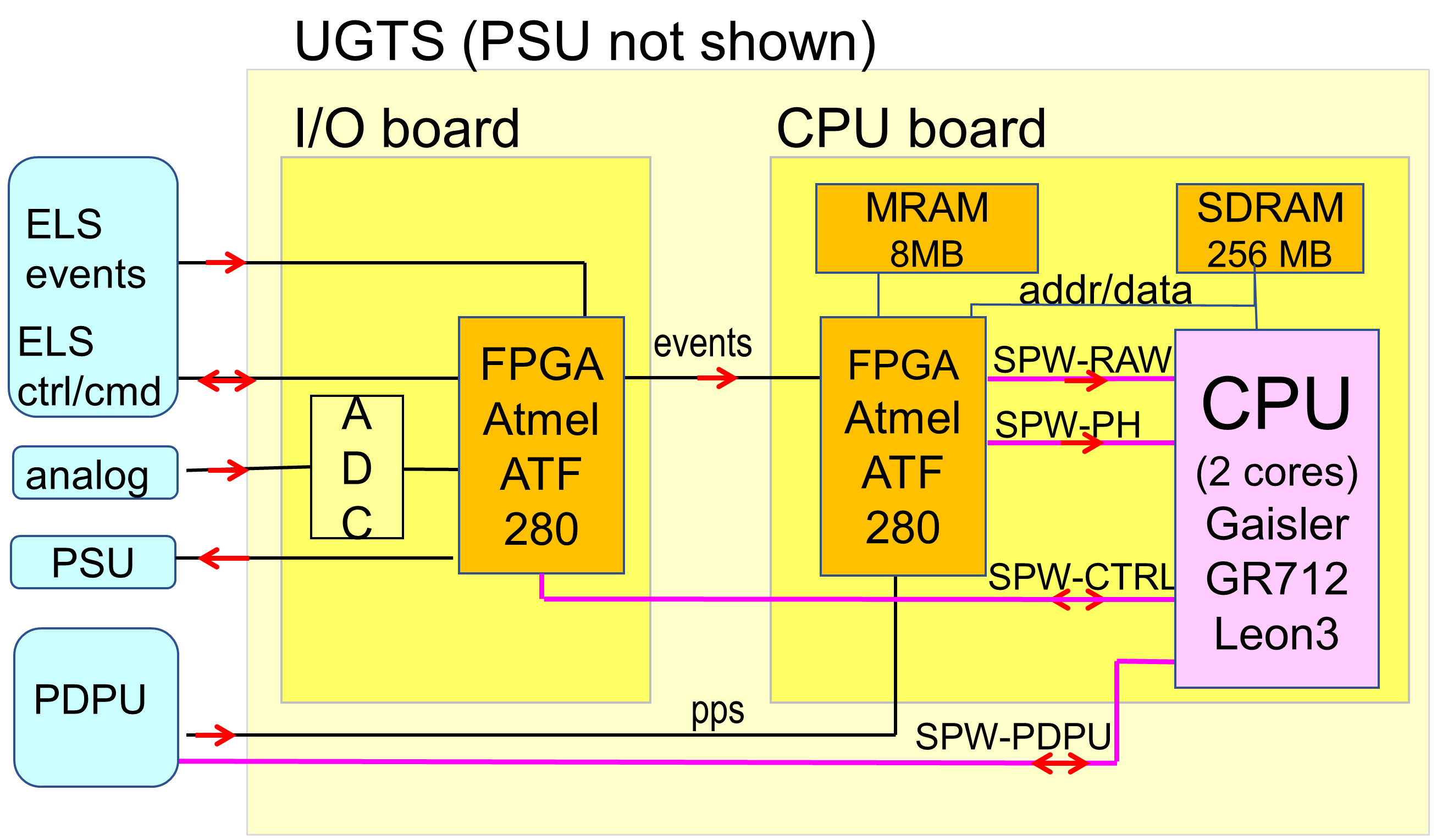}
		\caption{Electric architecture of UGTS (simplified), see text for details.}
		\label{fig_simplifiedElecArchiUgts}
	\end{figure}

	The UGTS controls ECLAIRs operations, manages the detector plane, acquires and stores detector data, performs onboard triggering and localization, generates VHF alert messages and issues autonomous slew requests.
	Its radiation-tolerant hardware (Fig.\ref{fig_simplifiedElecArchiUgts}), consists of power supply modules (PSU), Input/Output board (I/O) and Central-Processor Unit board (CPU), in a cold-redundant architecture.
	The I/O board manages the interface with the ELS and PSU through a Field-Programmable Gate-Array (FPGA). 
	The CPU board includes a second FPGA, managing the data acquisition and pre-formatting for the trigger, and a dual-core Leon3 CPU, running the OBSW based on a time-partitioned hypervisor, with most resources allocated to the trigger. The memory consists of a volatile SDRAM of 256 MB and a persistent MRAM of 8 MB, which holds the FPGA and OBSW binaries and configuration tables.
	The UGTS hardware was procured by CNES. The On-Board Software (OBSW, \citealt{IEEE_OBSW_Chateau}), including firmware, command-control, data acquisition, as well as the trigger software \citep{2019MmSAI..90..267S}, was developed by our team at CEA.
	
	\section{ECLAIRs Trigger design}
	
	\subsection{Requirements, Constraints and Choices}
	
	\subsubsection{Purpose}
	The ECLAIRs trigger analyzes photon data from the ELS in real time to detect and localize transient outbursts: (i) unknown sources classified as GRB candidates and (ii) outbursts from known cataloged X/$\gamma$-ray sources. Corresponding alerts are sent to the SVOM Payload Data Processing Unit (PDPU) for dissemination via the VHF network.
	The delay between the burst start time  (\emph{Tb}) and its localization (\emph{T0}) is typically $<$10 s for standard GRBs, depending on the source flux and evolution, but can be several minutes for long duration image triggers.
	
	\subsubsection{Output}
	The trigger outputs alert message sequences containing source detection and localization. For events above the \emph{AlertThreshold} (typically SNR$>$6.5$\sigma$), alerts are sent to ground, reaching the SVOM French Science Centre (FSC) within $\sim$15 s. For events above the \emph{SlewThreshold} (typically SNR$>$8$\sigma$ during commissioning and 7$\sigma$ afterward), an autonomous spacecraft slew is requested, usually completed within 2--3 min.
	All RawData (ELS photons), orbital and pointing data, scientific housekeeping logs (\emph{UssMessages}), and a copy of all VHF messages are stored in the PDPU mass-memory and later downlinked via X-band. These data are replayed on ground for trigger tuning and used for offline monitoring and analysis.
	
	\subsubsection{Imaging principle}
	Transient detection relies on frequent onboard coded-mask deconvolutions. Photon events are accumulated into detector \emph{shadowgrams} (80$\times$80 pixels) over selected energy bands and time windows, then deconvolved \citep{Goldwurm+etal+2022} into sky images\footnote{The sky image matrix is of size 200$\times$200 pixels for better FFT computation speed, the first row and column being non significant.}, providing counts \emph{SkyCnt}, variance \emph{SkyVar}, and SNR computed as \emph{SkySnr}=\emph{SkyCnt}/$\sqrt\emph{SkyVar}$.
	For each sky position, \emph{SkyCnt} is estimated from the difference between average counts in illuminated and obscured detector pixels. Sources appear as significant excesses in the SNR image. Sub-pixel localization is obtained by fitting a fixed-width 2D Gaussian Point-Spread Function (PSF), yielding $<$12 arcmin accuracy at threshold SNR=6.5. Deconvolution, the most CPU-intensive step, is FFT-based and requires $\sim$1 s per sky image on the UGTS hardware.
	The cosmic X-ray background (CXB) was estimated as dominant prior to launch. It produces a non-uniform detector illumination. For integration times $>$20 s, this component is modeled and removed by fitting a 2D quadratic polynomial to the shadowgram prior to deconvolution. 
	
	\subsubsection{Earth and orbital constraints}
	Due to the SVOM pointing strategy, the Earth is present in the ECLAIRs field of view about two-thirds of the time. Its position is computed onboard from orbital and attitude data received every second from the PDPU.
	Earth transits modulate the CXB, well corrected by the quadratic background model. 
	After launch, enhanced particle backgrounds near $\pm30^\circ$ latitude were observed; these are also well handled by the onboard background correction and deconvolution, fortunately with very few false alerts. 
	The trigger excludes sky regions occulted by the Earth, preventing false alerts originating from Earth. Terrestrial gamma-ray flashes (TGFs) are delegated to ground offline analysis.
	
	\subsubsection{The South-Atlantic Anomaly (SAA)}
	The SAA is managed through predefined orbital zones: SAA-core, SAA-ext, and outside SAA. In SAA-core, detector high voltage and trigger processing are disabled, the complete scientific sofware in reintialized when leaving SAA-core. 
	In the SAA-ext region, the trigger SNR thresholds may be modified (typically increased, depending on configuration). 
	An additional protection mechanism, Xoff, automatically disables event acquisition during periods of extreme count rates in SAA-ext, while still counting events and allowing the trigger to run transparently.
	During in-flight commissioning, the Xoff settings were defined such that when the multiple-event rate exceeds 1600 counts/s during 10 s, Xoff is activated (event data acquisition is disabled, the trigger receives no photon data, but continues to advance in time), and when this rate drops below 800 counts/s, event acquisition is re-enabled.
	
	\subsubsection{Alert types and onboard source catalog}
	Two alert types are produced: GRB candidates (GRB\_INIT) and catalog-source alerts (CAT). The onboard catalog stores known source positions, converted to local sky coordinates after each slew.
	
	GRB candidates are searched in Earth-free regions, away from known sources and field-of-view borders. Alerts are issued above \emph{AlertThresh}, and slews are requested above \emph{SlewThresh}. Thresholds are tuned in flight to maintain acceptable false alert and slew rates.
	
	Catalog sources trigger alerts when exceeding source-specific thresholds. To prevent repeated triggers, thresholds are temporarily raised after detection. Slews on CAT alerts are currently disabled, as ToO operations are sufficient.
	The catalog supports up to 256 sources (81 currently). For selected bright sources, modeled illumination patterns are fitted and subtracted from the shadowgram prior to deconvolution, enabling detection of weaker underlying transients; no CAT alerts are issued for such subtracted sources.
	
	\subsubsection{Constraints from GRB diversity}
\begin{figure}
		\centering
		\includegraphics[width=0.5\textwidth, angle=0]{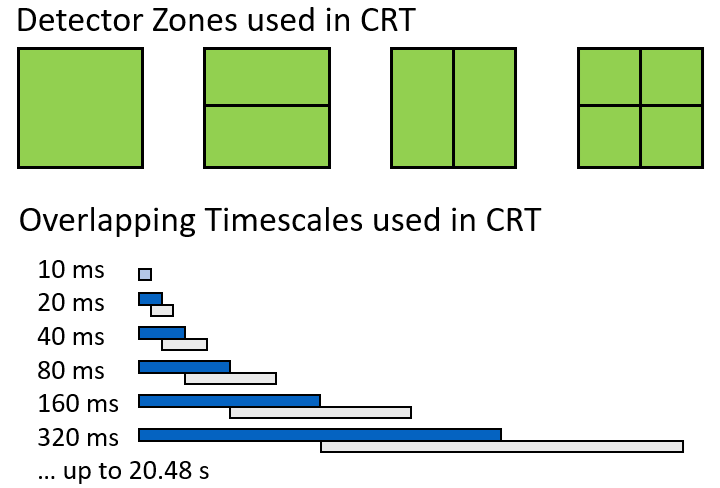}
		\caption{CRT searches for count-rate excesses over 9 detector zones, 4 Estrips, and overlapping timescales from 10 ms to 20.48 s.}
		\label{fig_crt-zones-and-timescales-scheme}
	\end{figure}

	GRBs occur randomly across the sky, requiring searches over nearly the entire field of view. Count-rates are evaluated on nine detector zones (Fig. \ref{fig_crt-zones-and-timescales-scheme}) to identify excesses.
	
	GRB fluxes and fluences span several orders of magnitude, while the faintest GRBs are the most numerous, requiring thresholds near the noise limit subject of producing false-alerts, to detect faint, distant events. 
	
\begin{figure}
		\centering
		\includegraphics[width=0.5\textwidth, angle=0]{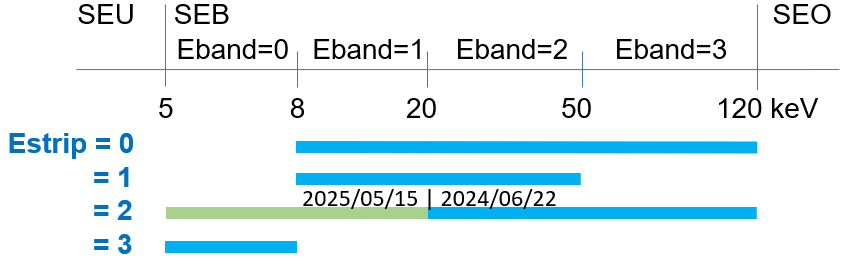}
		\caption{Mapping between adjacent Ebands encoded in single events (SEB) and overlapping Estrips used in the trigger. Estrip=2 was redefined during flight to improve GRB detection efficiency to soft GRBs.}
		\label{fig_energy-bands-and-strips}
	\end{figure}

	Spectral diversity is addressed by using four overlapping energy strips (Estrips), constructed from four adjacent energy bands (Ebands), configured in flight as 5--8, 8--20, 20--50, and 50--120 keV. Estrips are chosen to optimize sensitivity to soft as well as standard GRBs (Fig. \ref{fig_energy-bands-and-strips}).
	
	GRB durations range from milliseconds to tens of minutes. The trigger therefore analyzes time windows from 10 ms to $\sim$22 min, using overlapping time-windows to avoid sensitivity losses at window boundaries. With its long duration triggers, ECLAIRs shows an interesting discovery space for ultra-long GRBs \citep{2020ExA....50...91D}.
	
	\subsubsection{Constraints from the available CPU power}
	Since full sky deconvolution requires $\sim$1 s on the UGTS, it cannot be applied to all short time scales all the time. Two parallel algorithms are therefore implemented:
	\begin{enumerate}
		\item The Image Trigger (IMT), seeking GRBs on long time windows, performing in one step systematic sky reconstructions every 20 s, and summing them up to reach cumulative durations up to 22 min.
		\item The Count-Rate Trigger (CRT), seeking over all time windows from 10 ms to 20 s in two steps: first identifying count-rate excesses, and secondly reconstructing the sky image over the time window of the best excess identified.
	\end{enumerate}
	
	\subsection{Trigger algorithms}
	
\begin{figure}
		\centering
		\includegraphics[width=0.5\textwidth,
		angle=0]{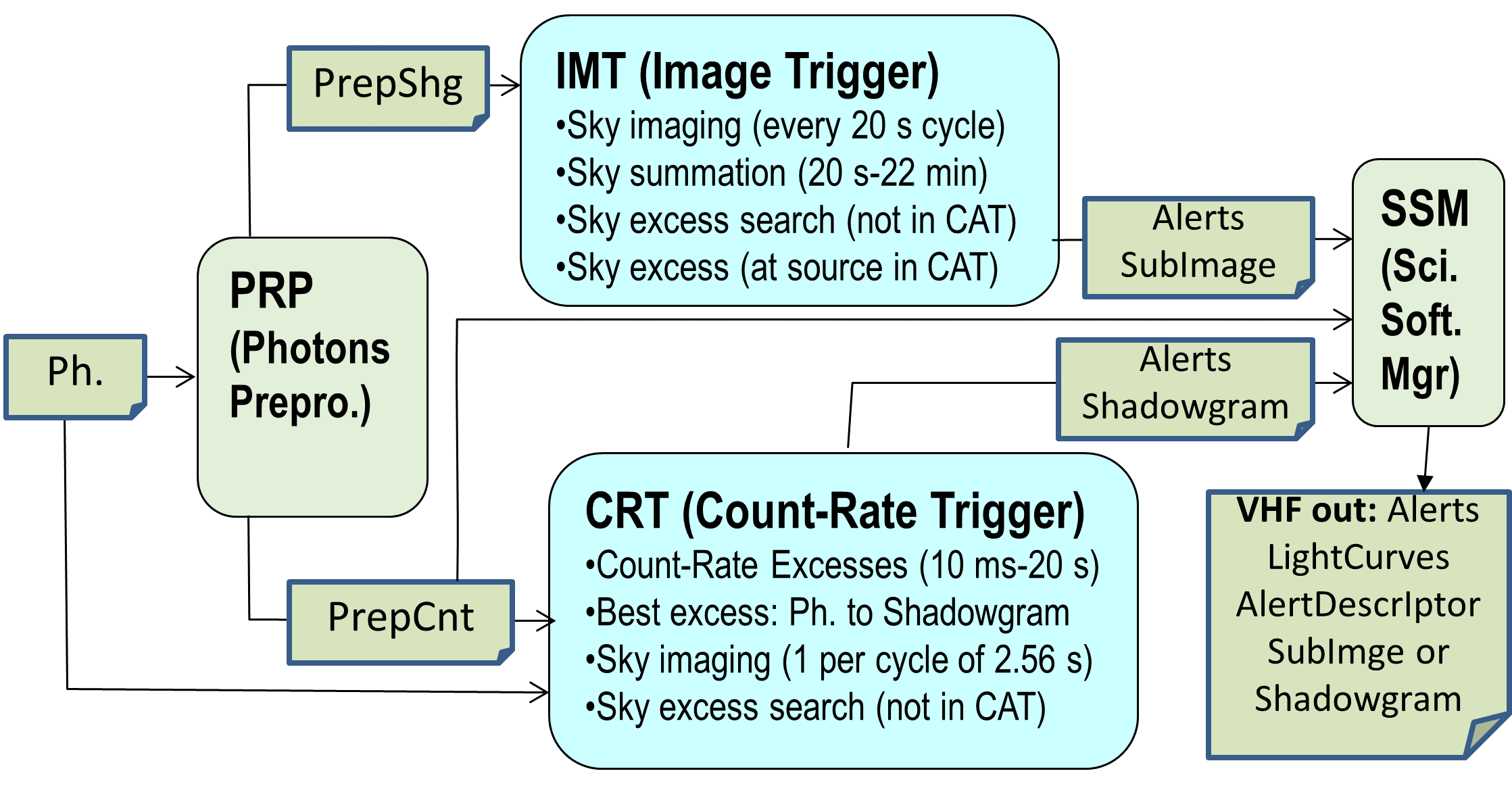}
		\caption{Schematics of the UGTS trigger algorithms and the interplay of the different software tasks involved.}
		\label{fig_USS_soft_interplay_details}
	\end{figure}

	The two trigger algorithms and their interplay in the UGTS software is shown on Fig. \ref{fig_USS_soft_interplay_details}.
	
	\subsubsection{Data pre-processing (PRP) for the triggers}
	
	The detector electronics ELS generate events of 32 bits, transmitted to the UGTS, synchronized by a \emph{TimeFrame} signal defining 10 ms intervals. The trigger uses only Single Events detected in Ebands (SEB). Other event types (under/overflows, multiples) are excluded from the trigger to reduce noise but retained for detector monitoring.
	
	An FPGA inside the UGTS manages event synchronization among all 8 ELS, buffering, and DMA transfers into \emph{RawData} packets stored in SDRAM and later forwarded to the PDPU for X-band downlink. This system supports event rates $>$100 kHz with minimal CPU usage.
	
	A reduced \emph{PhotonData} stream, containing only SEB events encoded on 16 bits, is built by the FPGA for trigger processing. \emph{TimeFrame} markers in the stream separate 10 ms periods, and counting information is also stored to permit global rate monitoring of high background conditions and activation of the Xoff protection mechanism.
	
	The trigger pre-processing task (PRP) running on CPU-core-0 scans the \emph{PhotonData} to build \emph{PrepShadowgrams} for IMT and \emph{PrepCounters} for CRT. 
	For computation efficiency, counters are stored incrementally in ring buffers, allowing rapid computation of counts over long durations by simple differencing rather than summing over many entries.
	Satellite orbital and attitude data are stored in the \emph{OrbitoData} buffer addressed by time.

	\subsubsection{Image Trigger (IMT) algorithm}
	
	The Image Trigger (IMT) runs on CPU-core-1 as a background task in an OBSW software partition, without strict real-time constraints. It processes \emph{PrepShadowgrams} sequentially from a ring buffer filled by the trigger PRP on CPU-core-0. This decoupling allows IMT to complete each cycle independently; if it falls behind real time, a fast-forward mechanism skips shadowgrams for triggering while preserving internal state. With a nominal cycle of 20.48 s (configurable down to 15 s), this mechanism is very rarely activated.
	
	{\bf \textbullet \ IMT-step1.0: initialization} At the start of each cycle, IMT checks for transitions: (i) presence of a GRM trigger activating temporary threshold reduction
	\footnote{In the case of a GRM trigger, the alert and slew thresholds in IMT and CRT can be temporarily lowered, depending on the configuration, potentially enabling the localization of a source exhibiting a long-lasting low-energy tail. In the default configuration, the alert threshold is reduced to SNR=6.0 for a few minutes when a GRM trigger involves three GRDs, possibly indicating a source within the ECLAIRs field of view. Such coincidences, however, remain rare with the current GRM trigger algorithm.}
	and (ii) occurrence of a spacecraft slew. 
	
	During a slew, IMT processing is suspended.
	After a slew or trigger restart (e.g. after SAA-core), sources from the onboard catalog within the new field of view are converted to local coordinates, exclusion masks are built around them, and illumination models for fitted sources are recomputed using current satellite attitude. Small attitude drifts without slew commands can be ignored, as platform stability is sufficient; a configurable margin added after slew allows for  stabilization delay if needed.
	
	IMT then computes the Earth-occulted sky region (with a configurable Earth limb margin) and identifies catalog sources hidden by the Earth, excluding them from fitting and CAT-trigger evaluation.
	
	{\bf \textbullet \ IMT-step1.1: sky image reconstruction} For each cycle, IMT reconstructs sky images at the base timescale (20.48 s by default) separately for each Estrip.
	To do so, first the shadowgram in given Estrip is built from the \emph{PrepShadowgrams}, cleaned from anomalously high-count pixels, and corrected using an efficiency map \citep{2024A&A...683A..60X}. A background remover then jointly fits illumination models of visible catalog sources and a 2D quadratic background, which are subtracted from the shadowgram. An alternative wavelet-based background removal method is implemented onboard \citep{2022A&A...665A..40D} but not yet activated in flight.
	
	The cleaned shadowgram is deconvolved into a sky image (counts and variance), using Estrip-dependent pixel weights to ignore dead or unstable pixels. Sky regions occulted by the Earth are set to zero to prevent noise accumulation in next step.
	
\begin{figure}
		\centering
		\includegraphics[width=0.5\textwidth, angle=0]{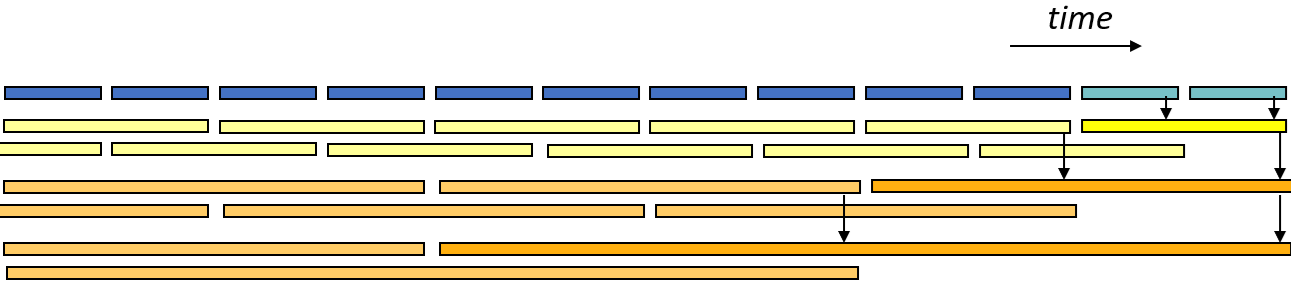}
		\caption{IMT sky-history summation scheme. Skies at Tscale=n are summed two by two to produce Tscale=n+1, resulting in two dyadic time-scale series. Each cycle, higher Tscales are computed when lower Tscales available.}
		\label{fig_imt-sky-summing-scheme}
	\end{figure}

	{\bf \textbullet \ IMT-step1.2: sky history construction} Each reconstructed sky at Tscale=0 is inserted into the sky-history buffer. Skies are summed pairwise to build longer integrations up to Tscale=6, covering durations from 20.48 s to 1310.72 s. A second set of half-shifted sums is also produced to preserve sensitivity to bursts near window boundaries. Higher Tscales are computed only when the required lower-level Tscales are available (Fig. \ref{fig_imt-sky-summing-scheme}).
	
	{\bf \textbullet \ IMT-step1.3: sky analysis and triggering} All newly built skies (for all Estrips and available Tscales) are analyzed. The SNR-image is computed, and the pixel with maximum SNR value \emph{SkySnrMax} is searched outside the catalog sources exclusion zones and poorly coded border region. Earth-masked regions having been set to 0 prior to summation are automatically taken into account. The SNR-image standard deviation \emph{SkySnrStd} and the second-highest SNR peak value \emph{SkySnrMax2} are also computed, providing robustness of the trigger condition against variable background.
	
	A GRB\_INIT alert is generated when all following conditions are met:
	\begin{enumerate}
		\item $\emph{SkySnrMax} > 6.5$
		\item $\emph{SkySnrMaxOvStd=SkySnrMax/SkySnrStd} > 6.5$
		\item $\emph{SkySnrDiffMax2=SkySnrMax-SkySnrMax2} > 1.5$.
	\end{enumerate}
	
	For alerts for which a spacecraft slew is proposed (SlewProposal=true), stricter thresholds apply (typically \emph{SkySnrMax} and \emph{SkySnrMaxOvStd}$>$8.5 during commissioning and $>$7 in nominal operations). Thresholds are configurable per Estrip, Tscale, orbital zone (e.g. SAA-ext), and GRM-trigger context.
	
	To avoid too numerous alerts, an IMT alert is issued if the current $\emph{SkySnrMax}$ exceeds the previous one (with comparison reset being configurable per alert sequence, Estrip, or even Tscale).
	
	Before transmission of the alert, the source peak is fit with a fixed-width 2D Gaussian to obtain sub-pixel localization. The 90\% confidence level (C.L.) error radius follows $R90 \simeq$ 77 arcmin/\emph{SkySnrMax} (plus $\sim$2 arcmin systematic). The fitted position is converted to local spacecraft and global (equatorial or galactic) coordinates; for Slew Requests, this position is used directly by the platform.
	
	IMT can also generate CAT alerts for catalog sources. For each sky image, the SNR at all catalog source positions is compared to source-specific thresholds (scaled by image duration). The most significant excess generates a CAT alert. To avoid repeated alert sequences on CAT sources, thresholds for triggered sources are raised at the end of an alert sequence.
	In the evaluation process of catalog-sources, their SNR values found in the sky images produced by IMT are recorded in \emph{UssMessages} for ground analysis.
	
	An \emph{AlertOnRequest} mode allows ground-commanded alerts at predefined positions (local or R.A., Dec.), used during commissioning to validate slews and alignment.
	
	\subsubsection{Count-Rate Trigger (CRT) algorithm}
	
	The Count-Rate Trigger (CRT) detects GRBs on short timescales (10 ms to 20.48 s), inaccessible to IMT due to CPU constraints. It runs sequentially to pre-processing (PRP) on CPU-core-0, and in parallel with IMT on CPU-core-1. CRT operates in 2.56 s cycles and performs at most one sky image reconstruction per cycle.
	
	Each cycle consists of two steps: (i) detection of count-rate excesses over multiple timescales and detector zones, stored in a ring buffer; (ii) imaging of the most significant recent excess to search for a new source.
	
	{\bf \textbullet \ CRT-step1.0: temporal background model update} CRT maintains 36 rolling background models (4 Estrips $\times$ 9 detector zones). Using incremental counters, up to 30 bins of 2.56 s are fit with a first-order polynomial after iterative outlier rejection. If valid, a second-order polynomial is then fitted and stored for background extrapolation.
	
	{\bf \textbullet \ CRT-step1.1: count-rate excess search} Newly available data are scanned over all configured timescales (23 by default, ranging from 10 ms to 20.48 s), including overlapping time windows, used to reduce trigger-sensitivity loss in case of unfavorable time phasing. For each time window, Estrip, and zone, background counts $B$ are extrapolated and compared to observed counts $N$ using a configurable SNR formula ($(N-B)/\sqrt{N+\emph{varMin}}$ applied, with $varMin$ protecting from counts close to 0). If the best zone exceeds the configured threshold (default 3$\sigma$), the excess is stored in the \emph{CountExcessRing}. Monitoring statistics are also recorded.
	
	{\bf \textbullet \ CRT-step2.0: extraction of best excess} The most significant non-imaged excess, not older than a configurable delay (default 10.24 s), is selected for imaging.
	
	{\bf \textbullet \ CRT-step2.1: detector plane image extraction} Photons corresponding to the selected time window and Estrip are extracted from the \emph{PhotonData} ring buffer and used to build the shadowgram.
	
	{\bf \textbullet \ CRT-step2.2: sky image reconstruction} The shadowgram is cleaned and efficiency-corrected, then deconvolved into a sky image. Unlike IMT, no spatial background or catalog-source fitting is applied. In flight, rare false alerts caused by particle backgrounds and strong sources (e.g. Sco X-1) can happen. Additional configuration safeguards include aborting the imaging step in CRT when Sco X-1 is visible (not behind Earth).
	Earth-occulted regions and catalog exclusion zones are identified using orbital and attitude data and masked in the sky image.
	
	{\bf \textbullet \ CRT-step2.3: sky image analysis and triggering} The sky is analyzed similarly to IMT. Alerts are generated when thresholds on $\emph{SkySnrMax}$, $\emph{SkySnrMaxOvStd}$, and $\emph{SkySnrDiffMax2}$ are met and depending on GRM-context. Issuing a SlewProposal requires passing higher thresholds. Thresholds are configurable per (\emph{Estrip}), time scale (\emph{Tscale}) and orbital conditions.
	
	CRT does not generate CAT alerts\footnote{Soft Gamma-ray Repeater sources are not included in the onboard CAT, such that CRT may trigger on their outbursts and issue a GRB-candidate alert. The case is then managed on ground by Trigger Advocates on shift and communicated to the community.}, to limit processing load, and as catalog-source flares on shorter timescales than for IMT are very unlikely. Nevertheless, SNR values at catalog-source positions found in CRT sky images are recorded in \emph{UssMessages} for ground analysis.

	\subsection{Trigger outputs}
	
	The GRB trigger running in the UGTS produces fixed-length CCSDS packets sent to the PDPU, either for real-time transmission via the VHF network or for delayed download via X-band.  
	VHF packets are buffered in prioritized FIFOs and transmitted as soon as possible, while X-band packets are stored in the PDPU mass memory and downloaded when a ground station is in visibility, making them unsuitable for ground analysis is real-time.
	
	\subsubsection{VHF Alert sequences}
	
\begin{figure}
		\centering
		\includegraphics[width=0.5\textwidth, angle=0]{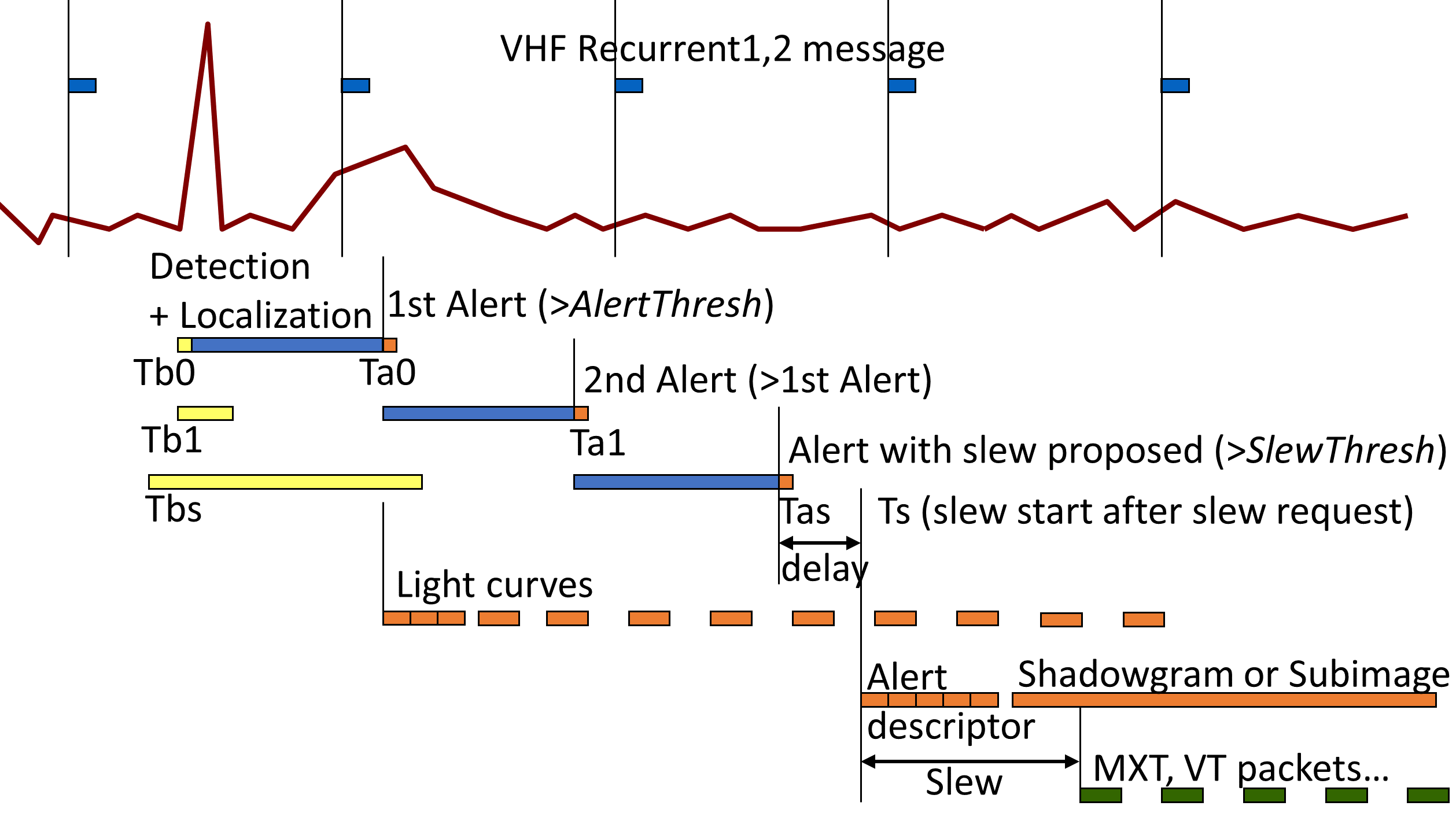}
		\caption{Timeline of a VHF Alert Sequence.}
		\label{fig_alert-sequence-scheme}
	\end{figure}

	When a trigger occurs, the UGTS initiates a \emph{VHF Alert Sequence} composed of packets of types: (i) VHF Alert, (ii) VHF Light Curve, (iii) VHF Alert Descriptor, and finally either (iv) VHF Subimage (IMT) or (v) VHF Shadowgram (CRT) (Fig.~\ref{fig_alert-sequence-scheme}).
	The packets are generated by the \emph{AlertService} within the Scientific Software Management (SSM) partition, which receives Alerts from IMT and CRT through independent FIFOs. Packets are interleaved, timestamped with a \emph{PacketTime} (the first Alert defines the trigger reference time) and sent-out at a maximum rate of 1 every 2 seconds (for Alerts).
	
	{\bf VHF Alerts}  
	Each Alert packet summarizes one sky-image detection, including: trigger origin and type, observation time window ($\emph{Tb}$, $\emph{Tscale}$), energy range ($\emph{Estrip}$), sky pixel and fitted sub-pixel position, local ($\Theta,\Phi$) and equatorial coordinates, SNR metrics ($\emph{SkySnrMax}$, $\emph{SkySnrStd}$, $\emph{SkySnrDiffMax2}$), spacecraft position and attitude, and a \emph{SlewProposal} flag.
	Upon receipt of the first Alert, the USSM starts a new AlertSequence and tracks the best detection (highest SNR). After a configurable delay (default 12 s), the best Alert defines the reference time for producing \emph{VhfLightCurve} packets.
	
	When an Alert packet contains a \emph{SlewProposal}, the best Alert packet after a configurable delay (default 26 s) is resent as with the \emph{SlewRequest} set. This delay allows multiple Alerts with higher SNR to be accumulated, improving confidence in the slew request and its localization accuracy. The platform then evaluates and either accepts or rejects the slew, based on its own constraints.
	If the slew is accepted, IMT and CRT stop triggering and the new source is added to the dynamic onboard catalog to prevent retriggering on it. If no slew is accepted, the AlertSequence terminates after a timeout (default 8 minutes).
	
	{\bf VHF Lightcurve packets}  
	The \emph{VhfLightCurve} consists of 64 packets providing count evolution in the 4 Ebands plus $Glo0$ (saturating) and $Glo1$ (multiples), for three binning resolutions: medium (0.8 s), low (6.4 s), and high (0.1 s). Each packet contains 4 time samples, the first being an incremental counter (enabling some recovery from packet losses). Spacecraft position and attitude are also included, allowing background evolution modeling during slews. For very long IMT timescales, light curves are not produced.
	
	{\bf VHF AlertDescriptor, Subimage and Shadowgram}  
	At the end of the sequence, \emph{AlertDescriptor} packets summarize the observation results, followed by the final \emph{VhfSubimage} (IMT) or \emph{VhfShadowgram} (CRT) corresponding to the best Alert.
	The \emph{VhfSubimage} consists of typically 49 packets covering a $(7 \times 8)^2$-pixel region around the trigger. Each packet contains a $8\times8$ pixel tile encoding SNR values with 8-bit quantization. Tiles are transmitted in a spiral pattern starting from the trigger pixel, prioritizing the most relevant region first.
	The \emph{VhfShadowgram}, produced only by CRT, is composed of 64 packets covering the full detector plane with $10\times10$ pixel tiles. Pixel counts are encoded on 6 bits (0--63), sufficient for counts integrated over the maximum CRT time window. On ground the sky image is reconstructed from this shadowgram, ignoring missing tiles.
	
	The VHF system is described in \citep{Cordier+etal+2026b}. It consists of 45 VHF receiver stations distributed along the satellite ground track. The median delay between sending of a VHF packet onboard, reception at a receiver and delivery to the FSC is $\sim$12 s.
	In some regions, such as over the Pacific Ocean, the network lacks receiver stations. Alert sequence messages are systematically repeated onboard for several tens of minutes (by the PDPU, with configurable settings) to reduce the probability of missed alerts, at the expense of a possible increase in delivery delay.
	
	\subsubsection{Monitoring messages}
	
	Trigger monitoring messages are either sent regularly in real-time over VHF or delayed over X-band.
	
	{\bf VHF Recurrent1}  
	sent every $\sim$60 s by IMT and CRT independently, these packets permit to monitor the trigger processing: 8 best sky excesses found last minute, strongest catalog source, background-fit information, and compact pointing data. 
	They are also processed in near-realtime on ground to produce \emph{SvomTriggerTargets} messages, enabling follow-up of possible transients below the Alert threshold by facilities allowing a larger false alarm rate, and in particular by GWAC for searching associated prompt optical emission when observing simultaneously the ECLAIRs field of view.
	
	{\bf VHF Recurrent2}  
	sent every 30 s by the OBSW, these packets report instrument mode, data acquisition status, spacecraft slews and SAA conditions, detector count rates, and key housekeeping parameters.
	
	{\bf USS messages}
	sent via X-band, those compact (10-byte) diagnostic messages provide detailed logs of trigger processing steps. Each message has a unique ID and a fixed data format, allowing unambiguous decoding on ground. They record quantities for each reconstructed sky-image by IMT and CRT such as SNR maxima, as well as catalog-source SNR evolutions, and are extensively used for offline analysis and trigger performance monitoring.

	
	\section{ECLAIRs Trigger implementation}
	
	The ECLAIRs Application Software was developed at CEA Paris-Saclay, with support from CNES, between 2016 and 2022 by a team of 15 people involved at different stages of the project. It includes the CPU-board FPGA firmware and OBSW \citep{IEEE_OBSW_Chateau}, as well as a wide range of test and validation tools. Early prototypes of the FPGA firmware \citep{Leprovost2013ieee} and the trigger software \citep{Schanne2013ieee} were implemented on the former Scientific Processing Unit (UTS) hardware developed at CEA, based on a Leon2 CPU, later replaced by the UGTS, based on a Leon3 CPU.
	
	{\bf UGTS On-Board SoftWare (OBSW)} 
	Most of the OBSW is written in C++17, while the more critical instrument control code is implemented in C. The software runs on the time-partitioning hypervisor XtratuM, organized into several so-called partitions that execute virtually in parallel and communicate through dedicated channels.
	
	{\bf UGTS Scientific Software (USS)} 
	The USS consists of about 100k lines of C++ code maintained in three GitLab projects: the basic components (\emph{ssbu}), the trigger code (\emph{sstu}), and the ground code (\emph{ssbg}). Both \emph{sstu} and \emph{ssbu} include the PRP, IMT, CRT, and SSM code. When used as OBSW submodules, it is compiled for the onboard Leon3 CPU into a single 607 KB binary, which can be updated in the ECLAIRs UGTS if required.
	The same binary is used for the USS (IMT, PRP/CRT and SSM) in separate partitions running in parallel, with different entry points. The objects of the USS code are redeployed in SDRAM each SAA-core exit, at most after 11 h of continuous operation, improving reliability.
	
	{\bf UGTS ground tests and simulations} 
	The same USS code can also be compiled on Linux systems as C++ libraries, interfaced with Python and imported as Python modules. A comprehensive set of simulation tools was developed and extensively used before launch, including a ray-tracing simulation of ECLAIRs for point-like and diffuse sources modulated by Earth passages. These tools were essential for USS development and testing in the absence of flight data, since the expected background variability could not be reproduced on the ground.
	During ground thermal-vacuum tests, the coded-mask imaging software (particularly in IMT) was validated using an X-ray generator and radioactive sources, by scaling the mask model to account for the finite source distance.
	
	{\bf UGTS local test unit (LTU)}
	The LTU ground test system was essential for the development and validation of the OBSW and firmware. It consists of a host computer that stores simulated RawData events representative of those produced by the ELS, and a photon injector that reads these events and injects them in real time—according to their time tags—into the UGTS hardware model, together with simulated orbital information provided every second. A large set of simulated datasets was produced to test and validate successive OBSW versions. OBSW version v7.0.3, selected for flight, remains the operational onboard version as of December 2025.
	The LTU is also used to replay flight RawData received via X-band, converted into representative ELS data, on a ground-based copy of the UGTS hardware. This capability is particularly useful for testing new software configurations with real flight data before deployment onboard.
	
	{\bf UGTS system validation bench (BVS)} 
	The BVS is a software emulator of the UGTS FPGA and CPU, running the same code as the UGTS hardware. It is used to test simulated or flight data with new software configurations or versions prior to onboard deployment. Multiple BVS instances can be run in parallel to speed up data reprocessing.
	
	\section{ECLAIRs Trigger in-flight performance}
	
	\begin{figure}
		\centering
		\includegraphics[width=0.55\textwidth]{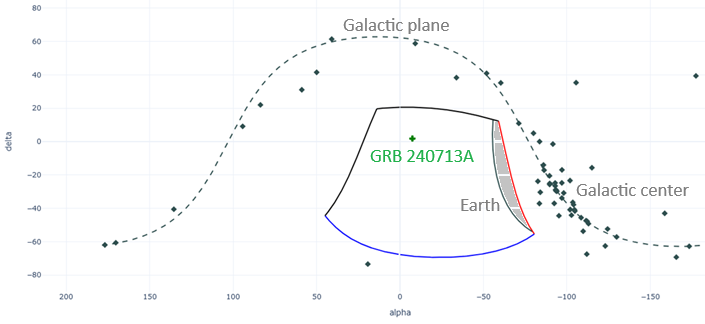}
		\caption{Trigger localization of GRB 240713A on the sky and withing the ECLAIRs field of view free of Earth.}
		\label{fig_GRB240713A_position-ra-dec}
	\end{figure}
	
	\begin{figure}
		\centering
		\includegraphics[width=0.35\textwidth]{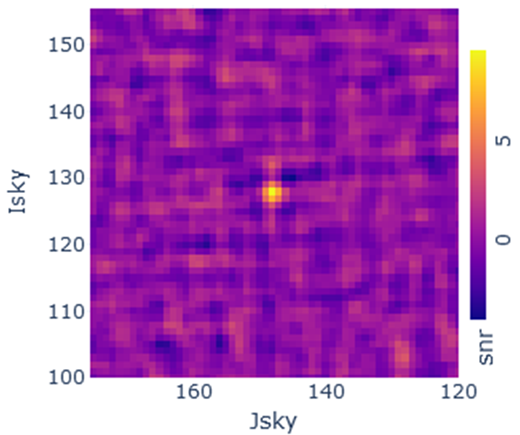}
		\caption{The point-like nature of the GRB 240713A trigger is confirmed by the SubImage sent by the IMT onboard trigger at the end of the VHF Alert Sequence (received in the X-band copy, since sending to VHF was still disabled).}
		\label{fig_GRB240713A_sub-image}
	\end{figure}
	
	Following the SVOM launch on June 22, 2024, the trigger was activated onboard ECLAIRs on July 11, and the first GRB was detected on July 13 by both the CRT and IMT, which produced three Alert packets each, recorded in the X-band duplication of the VHF packets since real-time VHF transmission was not yet activated (Figs. \ref{fig_GRB240713A_position-ra-dec} and \ref{fig_GRB240713A_sub-image}) \citep{2024GCN.36854....1S}. The event was confirmed as a sub-threshold GRB by Fermi/GBM.
	Since then, many notable bursts have been detected, including GRB 250314A, the most distant GRB observed in the past 12 years (z = 7.3) \citep{2025A&A...704L...7C}, which triggered an extensive follow-up campaign, including observations with JWST.
	
\begin{figure}
		\centering
		\includegraphics[width=0.5\textwidth]{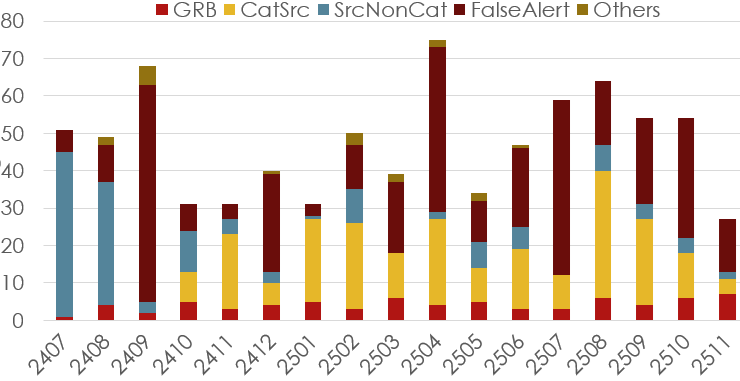}
		\caption{Monthly AlertSequences: GRB=confirmed GRBs, CatSrc=Catalog sources, SrcNonCat=Known sources not in catalog, FalseAlert=mostly due to detector plane inhomogeneities.}
		\label{fig_trigger-sequences-per-month}
	\end{figure}

\begin{figure}
		\centering
		\includegraphics[width=0.5\textwidth]{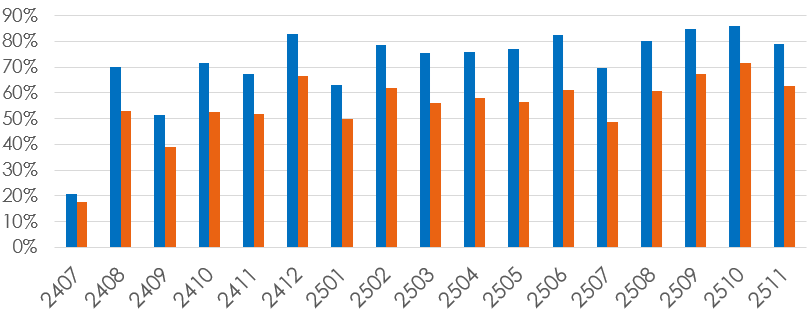}
		\caption{Monthly time fraction in operational mode (camera on, blue) and Xon (acquisition/trigger active, orange).}
		\label{fig_eclairs-oper-xon-per-month}
	\end{figure}

	From mid July 2024 to November 2025 (over 17 months), the ECLAIRs trigger produced 804 AlertSequences: 71 confirmed GRBs, 221 on CAT sources, 140 on known non-CAT sources, and 354 false alerts (Fig. \ref{fig_trigger-sequences-per-month}), yielding an average of 50.1 GRBs/year. False alerts mostly arise from inhomogeneous background on the detection plane, occasionally visible in the VHF subimage or shadowgram.
	The instrument was in operational mode (including tuning, camera on) on average 71.5\% of the time, including commissioning and deliberate downtime (Fig. \ref{fig_eclairs-oper-xon-per-month}). Peaks reach 86.0\% (October 2025), corresponding to the time outside the SAA-core. 
	
	However, high counting rates near $\pm$30$^o$ latitudes often activate Xoff (see details in \citealt{Claret+etal+2026}), where the data acquisition is switched off, preventing from triggering.
	Accounting for this, the Xon time fraction (trigger active) is 55.0\% on average (and 71.7\% in October 2025), with solar activity reducing observing conditions.
	Despite this reduced active-trigger time compared to pre-launch estimates, the confirmed GRB rate is 50.1/year. In the future under more quiet solar conditions, reaching 85\% Xon time could increase the GRB trigger rate to about 76/year.
	
	\section{Conclusions}
	
	
	Following early concept studies conducted during the ECLAIRs microsatellite project \citep{2005ITNS...52.2778S}, the trigger algorithm designs \citep{2008ICRC....3.1147S} and prototypes \cite{Schanne2013ieee} introduced several innovative concepts implemented onboard, which we are delighted to operate successfully in flight since 1.5 years.
	
	
	Among these concepts, crucial for achieving clean sky reconstructions are the removal of multiple events at detector/data-acquisition level, the cleaning of abnormal/noisy pixels in each shadowgram and the spatial background and known-source subtraction (for 20 s exposures) prior to deconvolution.
	For CRT, the main innovation lies in its two-step algorithm, carefully selecting excesses prior to imaging. The first step uses an elaborate sliding temporal background model based on time-series fitting to identify count-rate excesses. In the second step, these excesses are ranked by score and only the most significant one is selected for imaging.
	For IMT, the key innovation enabling triggering on very long exposures (22 min, previously unattained) is the background and known-source subtraction performed each 20 s cycle at the shadowgram level to correct inhomogeneities variable at those timescales, including Earth passages, followed by the summation of cleaned sky images to achieve ultra-long exposures.
	
	The ECLAIRs onboard trigger performs remarkably well in detecting GRBs for SVOM, triggering in the so far unexplored low energy ($>$5 keV) domain, while maintaining a relatively low false-alert rate despite an in-flight background that has proven far more variable than anticipated, owing to enhanced particle fluxes along the orbit during the current phase of high solar activity. Further configuration tuning and software updates are planned to enable triggering with lower thresholds, thereby detecting even fainter (potentially high-redshift) GRBs in the future.
	
	\begin{acknowledgements}
		The Space-based multi-band astronomical Variable Objects Monitor (SVOM) is a joint Chinese-French mission led by the Chinese National Space Administration (CNSA), the French Space Agency (CNES), and the Chinese Academy of Sciences (CAS). 
		We acknowledge the unwavering support of NSSC, IAMCAS, XIOPM, NAOC, IHEP, CNES, CEA, and CNRS.
		The authors would like to thank for their valuable contributions at various stages of the ECLAIRs onboard software development: S. Anvar, C.-H. Besson, I. Breschi, E. Delagnes, Ch. Flouzat, Ch. Hervioux, L. Klenov, K. Metzher, P.-F. Rocci, C. Tahoulan, A. Valiente, N. van Hille, as well as the CNES team for their support: M.-C. Charmeau, J. Galizzi, F. Gonzalez, Ph. Guillemot, L. Perraud.
		The SVOM/ECLAIRs trigger scientist (corresp. author) warmly thanks members from the collaborations of Swift/BAT (E. Fenimore, D. Palmer, M. Galassi, A. Lien, S. Barthelmy),  INTEGRAL/IBAS (S. Mereghetti, D. Götz), Fermi/GBM (A. von Kielnin) and the ECLAIRs team including students (J.-L. Atteia, B. Cordier, A. Gros, A. Goldwurm, O. Barrière, B. L'Huillier, M. Courtois, M. Cortial, D. Zhao, S. Antier, W. Xie) for fruitful discussions.
	\end{acknowledgements}
	

	\label{lastpage}
	
\end{document}